\documentclass[reqno]{amsart}
\usepackage{url}
\usepackage[plainpages=false,pdfpagelabels,colorlinks=true,citecolor=blue,hypertexnames=false]{hyperref}

\def\erf{\operatorname{erf}}
\newtheorem{theorem}{Theorem}
\newtheorem{lemma}{Lemma}

\newtheorem{prop}{Proposition}

\theoremstyle{definition}
\newtheorem{ex}{Example}

\begin{document}
\title{A Proof of a Recursion for Bessel Moments}

\author{Jonathan M. Borwein}
\thanks{Research supported in part by NSERC and the Canada Research Chair Programme.}
\address{Faculty of Computer Science,
Dalhousie University,
Halifax, NS, B3H 2W5,
Canada}
\email{jborwein@cs.dal.ca}

\author{Bruno Salvy}
\thanks{Research supported in part by the French \emph{Agence Nationale pour la Recherche} (ANR Gecko) and the Joint Inria-Microsoft Research Centre.}
\address{Algorithms Project,
Inria Paris-Rocquencourt,
78153 Le Chesnay Cedex,
France}
\email{Bruno.Salvy@inria.fr}
\date{February 28, 2008}
\maketitle

\begin{abstract} We provide a proof of a conjecture in \cite{BaBoBoCr07} on the existence and form of linear recursions for moments of powers of the Bessel function~$K_0$.
\end{abstract}

\section{Introduction}
 The aim of this  note is two-fold. First, we prove    a     conjecture    of~\cite{BaBoCr06, BaBoBoCr07}     concerning   the existence of  a    recurrence in $k \ge 0$     satisfied    by     the    integrals
\[C_{n,k}:=\frac1{n!}\int_0^\infty\dots\int_0^\infty\frac{dx_1dx_2\cdots  dx_n}{(\cosh   x_1+\cdots+\cosh  x_n)^{k+1}},\]
for $n=1,2,\ldots$. These integrals naturally arose during  the analysis of parts of the Ising theory of solid-state physics \cite{BaBoCr06}.  In~\cite{BaBoBoCr07} only the first four cases of
 Theorem \ref{main} below were proven and the proofs relied on the ability to express the corresponding integrals in (\ref{integral}) below as \emph{Meijer G-functions}, something which fails for $n>4$.

A second aim is to advertise the power of current symbolic computational tools and related algorithmic developments to settle such questions.    For this reason we give quite detailed exposition of the methods entailed.

 Our main
result (Theorem \ref{main}) is better  phrased in terms of  \begin{equation}\label{bigC} c_{n,k}:=\frac{n!\Gamma(k+1)}{2^n}C_{n,k}. \end{equation}

\begin{theorem}[Linear Recursion]\label{main} For any fixed $n\in\mathbb{N}$, the sequence~$c_{n,k}$ enjoys a linear recurrence with polynomial coefficients
of  the   form
\begin{equation}\label{main-eq}(k+1)^{n+1}c_{n,k}+\sum_{\substack{2\le  j<n\\  \text{$j$   even}}}{P_{n,j}(k)c_{n,k+j}}=0,
\end{equation}
with  $\deg P_{n,j}\le    n+1-j$.
\end{theorem}
\noindent   Substituting~\eqref{bigC}    and    simplifying    by~$n!\Gamma(k+1)/2^n$
yields
\[(k+1)^{n+1}C_{n,k}+\sum_{\substack{2\le  j<n\\  \text{$j$   even}}}{P_{n,j}(k)(k+j)(k+j-1)\dotsm(k+1)C_{n,k+j}}=0,
\]
which is~\cite[Conjecture~1]{BaBoBoCr07}, with extra information added on the  origin of the linear factors  for the recurrence in~$C_{n,k}$.


The starting point of our proof is the integral representation~\cite[Eq.~(8)]{BaBoBoCr07}
\begin{equation}\label{integral}
c_{n,k}=\int_0^{+\infty}{t^kK_0(t)^n\,dt},
\end{equation}
where~$K_0$ is the \emph{modified Bessel function} on which much information is to be found in~\cite[Ch.~9]{AbSt73}.
  The key  properties of $K_0$ that we use are as follows. First
  $$K_0(t)=\int_0^\infty e^{-t \,{\rm cosh}(x)}\,dx$$ which explains how the integrals in (\ref{integral}) arise. Moreover, we have
\begin{itemize}
\item[--] a linear differential equation: $(\theta^2-t^2)K_0(t)=0$ with~$\theta:=t d/dt$;
\item[--]  the behaviour at the origin: $K_0(t)\sim-\ln t$, $t\rightarrow0$;
\item[--] and behaviour at infinity: $K_0(t)\sim\sqrt{\pi/2t}e^{-t}$, $t\rightarrow+\infty.$
\end{itemize}
These last two properties show that the integral~\eqref{integral} converges for any complex~$k$ subject to~$\Re k>-1$ where it defines an analytic function of~$k$. The recurrence of Theorem~\ref{main} then gives the integral a meromorphic continuation to the whole complex plane with poles at the negative integers.

\section{Existence of a Recurrence}
The theory of D-finite functions leads to a direct proof of the existence of a recurrence such as~\eqref{main-eq} in a very general setting, together with an algorithm.

Recall that a function is called \emph{D-finite} when it satisfies a linear differential equation with polynomial coefficients. A good introduction to the basic properties of these functions is given in~\cite{Stanley99}. What makes these functions appealing from the algorithmic point of view is that they live in finite-dimensional vector spaces and thus many of their properties can be explicitly computed by linear algebra in finite dimensions. In this setting, the following proposition is easily obtained. It is a generalization of our main Theorem \ref{main}, except for the absence of degree bounds.

\begin{prop}\label{stanley} Assume that $f(t)$ obeys a homogeneous linear differential equation
\[a_r(t)f^{(r)}(t)+\dots+a_0(t)f(t)=0,\]
with polynomial coefficients~$a_i(t)$
in~${\mathbb C}[t]$.
For a fixed~$n\in{\mathbb N}\setminus\{0\}$, let $\Gamma$ be a path in~$\mathbb C$ such that for any~$k,j\in {\mathbb N}$ the integrals
\[I_{k,j}:=\int_{\Gamma}{t^k(f(t)^n)^{(j)}\,dt}\]
converge and the limits of the integrand at both endpoints coincide. Then  the sequence~$\{I_{k,0}\}$ obeys a linear recurrence with coefficients that are polynomial in~$n$ and~$k$ and which can be computed given the coefficients~$a_i$'s.
\end{prop}
We give the proof in two steps. The first one is classical and can be found for instance in~\cite[Thm.6.4.9]{Stanley99}.
\begin{lemma}
D-finite functions form an algebra over the rational functions.
\end{lemma}
This means that any polynomial in D-finite functions with rational function coefficients defines a functions that is itself D-finite. In particular~$K_0^n$ satisfies a linear differential equation.

\begin{proof}
The proof is effective. The difficult part is the product. The derivatives of two D-finite functions~$f$ and $g$ live in finite-dimensional vector spaces generated by~$f,\dots,f^{(r-1)}$ and $g,\dots,g^{(s-1)}$. Therefore by repeated differentiation the derivatives of a product~$h:=fg$ can be rewritten as linear combinations of the terms~$f^{(i)}g^{(j)}$, $0\le i<r$, $0\le j<s$ which generate a vector space of dimension at most~$rs$. It follows that the $rs+1$ successive~$h^{(k)}$, $k=0,\dots,rs$ are linearly dependent. A linear dependency between them can be found as the kernel of the linear map~$(\lambda_0,\dots,\lambda_{rs})\mapsto\lambda_0h+\dots+\lambda_{rs}h^{(rs)}$. Any such linear dependency is a linear differential operator annihilating~$fg$.
\end{proof}

The corresponding algorithm is implemented, among other places, in the \emph{Maple} package {\tt gfun}~\cite{SaZi94}.
\begin{ex}\label{ex1}Here is how the function {\tt gfun[poltodiffeq]} is invoked to compute a differential equation for~$K_0^4$:
\begin{verbatim}
> eqK0:=t*diff(t*diff(y(t),t),t)-t^2*y(t);
\end{verbatim}
\vskip-1ex
\[eqK0:=t (t\frac{d^2}{dt^2}y( t)+ {\frac{d}{dt}}y ( t )) -{t}^{2}y( t) \]	
\begin{verbatim}
> gfun[poltodiffeq](y(t)^4,[eqK0],[y(t)],y(t))=0;
\end{verbatim}
\vskip-3ex
\begin{multline*}
  t^{4} {\frac {d^{5}}{dt^{5}}}y ( t )
+10t^{3}{\frac {d^{4}}{dt^{4}}}y ( t )
- (20t^{4} - 25t^{2}) {\frac {d^{3}}{dt^{3}}}y ( t ) \\
- ( 120t^{3} -15t) {\frac {d^{2}}{dt^{2}}}y ( t )
+ (64t^{4} -152t^{2}+1 ) {\frac {d}{dt}}y ( t )
+(128t^{3} -32t ) y ( t ) =0
\end{multline*}
\end{ex}
\begin{ex}
Here are the corresponding steps of the calculation for the smaller example~$K_0^2$:
\begin{multline*}
h=K_0^2,\quad
h'=2K_0K_0',\quad
h''=2K_0'^2-2t^{-1}K_0K_0'+2K_0^2,\\
h^{(3)}=-6t^{-1}K_0'^2+4(2+t^{-2})K_0K_0'-2t^{-1}K_0^2,
\end{multline*}
where whenever possible we have replaced~$K_0''$ by $K_0-t^{-1}K_0'$.
Then we find the vector $(-4t,1-4t^2,3t,t^2)$
in the kernel of
\[\begin{pmatrix}1&0&2&-2t^{-1}\\ 0&2&-2t^{-1}&4(2+t^{-2})\\ 0&0&2&-6t^{-1}\end{pmatrix}.\]
This  vector then produces a differential equation satisfied by~$K_0^2$:
\[t^2y^{(3)}+3ty''+(1-4t^2)y'-4ty=0.\]
\end{ex}

\begin{proof} (continued)
The second step of the proof of  Proposition \ref{stanley} starts by expanding the differential equation for~$h:=f^n$ as
\[\sum_{i,j}d_{i,j}t^ih^{(j)}=0,\]
for scalars $d_{i,j}$.
This is then multiplied by~$t^k$ and integrated along~$\Gamma$. Use of the convergence hypotheses then allows us to deduce that
\begin{equation}\label{sumofint}
\sum_{i,j}d_{i,j}\int_{\Gamma}{t^{k+i}h^{(j)}\,dt}=0.
\end{equation}
Now, integration by parts gives
\begin{align*}
\int_{\Gamma}{t^{k+i}h^{(j)}\,dt}&=\underbrace{\left.t^{k+i}h^{(j-1)}\right|_{\Gamma}}_{0}-(k+i)\int_{\Gamma}{t^{k+i-1}h^{(j-1)}\,dt},\\
&=(-k-i)(-k-i+1)\dotsm(-k-i+j-1)I_{k+i-j},
\end{align*}
the last equality following by induction. Adding the contributions of all the terms in~\eqref{sumofint} finally yields the desired recurrence over~$I_k$. \end{proof}

\begin{ex}For~$h:=K_0^2$, the computation gives
\[\int_0^{+\infty}{t^{k+2}h^{(3)}+3t^{k+1}h''+(t^k-4t^{k+2})h'-4t^{k+1}h\,dt}=0,\]
whence
\begin{multline*}
(-k-2)(-k-1)(-k)c_{2,k-1}+3(-k-1)(-k)c_{2,k-1}\\
+(-k)c_{2,k-1}-4(-k-2)c_{2,k+1}-4c_{2,k+1}=0.
\end{multline*}
Once simplified, this reduces to
\begin{equation}\label{recc2}
4(k+1)c_{2,k+1}=k^3c_{2,k-1}.
\end{equation}
\end{ex}
\begin{ex} Quantum field theorist David Broadhurst has recently studied~\cite{Broadhurst07} the \emph{vacuum-diagram} integrals for $n\ge 0$:
\[ \label{vac} V(n,a,b):= \int_0^\infty x^{2n+1} K_0^a(x)(xK_0'(x))^b\,dx\]
and provides the recursion
\[2(n+1)V(n,a,b)+a\, V(n,a-1,b+1)+b\,V(n+1,a+1,b-1)=0\]
which \emph{preserves} $N:=a+b>0$ and allows one to reduce to $V$ values with $ab=0$. Note that $K_0^{'}=-K_1$.
Proposition \ref{stanley} applies to $n \mapsto V(n,a,b)$ for each $a$ and $b$ and as in Section \ref{alg} below leads to very efficient code for the recursion. The difficult question of understanding the initial values is discussed in~\cite{Broadhurst07} and~\cite{B3G08}.
\end{ex}

\paragraph{\bf Mellin transform}
As the proof indicates,  Proposition \ref{stanley} is not restricted to integer values of~$k$. In particular, the method gives a difference equation for the \emph{Mellin transform}
\[h^\star(s):=\int_0^{+\infty}t^{s-1}h(t)\,dt\]
provided the appropriate convergence properties are satisfied. This difference equation then gives a meromorphic continuation in the whole complex plane. The most basic example is~$\Gamma(s)$: starting from the elementary differential equation~$y'+y=0$ for~$h(t)=\exp(-t)$ leads to the classical functional equation~$\Gamma(s+1)=s\Gamma(s)$.

\paragraph{\bf Coefficients}
The path~$\Gamma$ can also be a closed contour. For instance, if~$h$ is analytic at the origin, then the $k$th coefficient of its Taylor series at the origin is given by the Cauchy integral
\[\frac 1{2\pi i}\oint{\frac{h(t)}{t^{k+1}}\,dt},\]
where the contour encloses the origin and no other singularity of~$h$. The algebraic manipulations are the same as in the previous case, followed by replacing~$k$ by~$-k-1$ and the sequence~$c_{k}$ by the sequence $u_{-k-1}$.

For instance, if we apply this transform to the functional equation for~$\Gamma$, we get
first $c_{-s}=-sc_{-s-1}$ and then
$u_{s-1}=-su_{s}$,
which is the expected recurrence
for the sequence of coefficients~$u_s=(-1)^s/s!$ of~$\exp(-t)$.

Similarly, starting from~\eqref{recc2} we obtain the mirror recurrence
\[4k\tilde{c}_{k-1}=(k+1)^3\tilde{c}_{k+1}.\]
Observe that this is obeyed by the coefficients of~$\ln^2(t)$ in the series expansion
\begin{eqnarray*}
K_0^2(t)&=&\ln^2(t)\left(1+\frac{1}{2}t^2+\frac{3}{32}t^4+\frac{5}{576}t^6+\dots\right)\\
&+&\ln(t)\left(2\gamma-2\ln 2+(\gamma-\ln 2-\frac{1}{2})t^2+\dots\right)
+\left((\ln 2-\gamma)^2+\dots\right),\quad t\rightarrow 0^+.
\end{eqnarray*}
The Frobenius computation of expansions of solutions of linear differential equations at regular singular points~(see, e.g.,~\cite{Ince56}) explains why this is so.

\paragraph{\bf Further generalizations}
The sequence $\{t^k\}$ in the integral of the proposition can be replaced by more general sequences of functions that satisfy both a linear recurrence in~$k$ and a linear differential equation in~$t$. Provided that proper analytic  conditions are satisfied at the endpoints of the path, the same result will hold. This makes it possible to compute, for instance, recurrences for the Fourier coefficients with respect to various bases such as orthogonal polynomials, Bessel functions,\dots.

The algorithm that applies in this case is called \emph{creative telescoping} as discovered by
Zeilberger~\cite{Zeilberger90} and further automated in~\cite{ChSa98,Chyzak00}. Again, the computation boils down to linear algebra in a suitably constructed finite-dimensional vector space.

In summary, all these algorithms succeed in making effective and efficient the familiar method of differentiation under the integral sign and integration by parts.

\section{Proof of the Main Result}
If $A$ is a linear differential operator,
the operator of minimal-order annihilating the $n$th power of every solution of~$A$ is called
its $n$th \emph{symmetric power}. Because of its role in algorithms for differential Galois theory~\cite{vdPSi02} there has been interest in efficient algorithms computing symmetric powers. In the case of \emph{second order} operators, such an algorithm has been found in~\cite{BrMuWe97}. We state it in terms of the derivation~$\theta:=td/dt$ in order to get better control over the coefficients of the resulting recurrence---but the statement and proof hold for any derivation.
\begin{lemma}[Linear Differential Equation, \cite{BrMuWe97}]\label{lde}
Let~$A=\theta^2+a(t)\theta+b(t)$ be a linear differential operator with rational function coefficients~$a(t)$ and $b(t)$. Let $L_0=1$, $L_1=\theta$ and for $k=1,2,\dots, n$ define the operator~$L_{k+1}$ by
\begin{equation}\label{defLk}
	L_{k+1}:=(\theta+ka)L_k+bk(n-k+1)L_{k-1}.
\end{equation}
Then, for $k=0,\dots,n+1$, and for an arbitrary solution~$y$ of~$Ay=0$,
\[L_ky^n=n(n-1)\dotsm(n-k+1)y^{n-k}(\theta y)^k\]
and in particular $L_{n+1}y^n=0.$
\end{lemma}
[This recursion can be viewed as an efficient computation of the kernel which was described in the previous section, taking advantage of the special structure of the current matrix.]

\begin{proof}
The proof is a direct verification by induction. For $k=0$ and $k=1$ the identity reduces respectively to~$y^n=y^n$ and $\theta y^n=ny^{n-1}\theta(y)$ which are obviously true for any function~$y$.
Assuming the identity to hold up to~$k\ge1$, the heart of the induction is the rule for differentiation of a product $\theta(uv)=\theta(u)v+u\theta(v)$:
\begin{align*}
	\theta(y^{n-k}(\theta y)^k)
	&=\theta(y^{n-k})(\theta y)^k+y^{n-k}\theta((\theta y)^k)\\
	&=(n-k)y^{n-k-1}(\theta y)^{k+1}+ky^{n-k}(\theta y)^{k-1}(\theta^2 y)\\
	&=(n-k)y^{n-k-1}(\theta y)^{k+1}+ky^{n-k}(\theta y)^{k-1}(a \theta y+by).
\end{align*}
Reorganizing terms concludes the induction.
\end{proof}

\begin{ex}\label{four} In the case of~$K_0$, we have $a=0$ and $b=-t^2$. For $n=4$, starting with~${L}_0=1$ and ${L}_1=\theta$,  the recurrence of Lemma~\ref{lde} gives
	\begin{align*}
	{L}_2&=\theta^2-4t^2,\\
	{L}_3&=\theta^3-10t^2\theta-8t^2,\\
	{L}_4&=\theta^4-16t^2\theta^2-28t^2\theta+8t^2(3t^2-2),\\
	{L}_5&=\theta^5-20t^2\theta^3-60t^2\theta^2+8t^2(8t^2-9)\theta+32t^2(4t^2-1).
	\end{align*}
The operator $L_5$ annihilates $K_0^4$. It is a rewriting in terms of~$\theta$ of the equation of Example~\ref{ex1}.
\end{ex}
\noindent Some of the patterns that emerge on this example can be proved in the general case.
\begin{lemma}[Closed Form]\label{lem2} With the same notation as in Lemma \ref{lde}, when $A=\theta^2-t^2$, $L_k$ may be written as
	\[L_k=\theta^k+\sum_{j=0}^{k-2}a_j^{(k)}(t)\theta^j,\]
	where each $a_j^{(k)}$ is a polynomial in~$t^2$, divisible by~$t^2$  and~$\deg a_j^{(k)}\le k-j$.
\end{lemma}
\begin{proof}
Again, the proof is by induction. For $k=0$ and $k=1$ we recover the definition of~$L_0$ and $L_1$. For larger~$k$, the recurrence~\eqref{defLk} simplifies to
\[L_{k+1}:=\theta L_k-k(n-k+1)t^2L_{k-1}.\]
If the property holds up to~$k\ge1$, then this shows that the degree of~$L_{k+1}$ in~$\theta$ is~$k+1$, with leading coefficient~1 and also that the coefficient of~$\theta^k$ in~$L_{k+1}$ is~0. Extracting the coefficient of~$\theta^j$ then gives
\[a_j^{(k+1)}=\begin{cases}a_{j-1}^{(k)}+\theta(a_j^{(k)})-k(n-k+1)t^2a_j^{(k-1)},&\qquad 0\le j\le k-2,\\
-k(n-k+1)t^2,&\qquad j=k-1.
\end{cases}\]
These last two identities give the desired degree bound and divisibility property for the coefficients~$a_{j}^{(k+1)}$, $0\le j\le k-1$.
\end{proof}

We may now complete the proof of the main result.

\begin{proof} (of Theorem \ref{main}.) Lemma \ref{lem2} shows that the coefficients of~$L_{n+1}$ can be rewritten as
\begin{equation}\label{finaldeq}
L_{n+1}=\theta^{n+1}+\sum_{\substack{2\le j<n\\ \text{$j$ even}}}{t^jQ_j(\theta)},
\end{equation}
where the polynomials~$Q_j$ satisfy $\deg Q_j\le n+1-j$.

Thanks to the properties of~$K_0$ recalled in the Introduction, an integration by parts yields
\begin{equation}\label{intparts}
\int_0^{+\infty}{t^{k+j}\theta^m(K_0^n(t))\,dt}=(-1-k-j)^mc_{n,k+j},
\end{equation}
for each $m$. Now, we multiply $L_{n+1}K_0^n$ from~\eqref{finaldeq} by~$t^k$ and integrate from~0 to infinity:
\[\int_0^{\infty}{\bigl\{t^{k}\theta^{n+1}K_0^n(t)+\sum_{\substack{2\le j<n\\ \text{$j$ even}}}{t^{k+j}Q_j(\theta)}K_0^n(t)\bigr\}\,dt}=0.\]
Integrating term by term and using~\eqref{intparts} finally gives the recurrence
\[(-k-1)^{n+1}c_{n,k}+\sum_{\substack{2\le j<n\\ \text{$j$ even}}}{Q_j(-1-k-j)c_{n,k+j}}=0,\]
which is the desired one, up to renaming and sign changes.
\end{proof}

\section{Algorithm}\label{alg}
In summary, we have a relatively straightforward algorithm to compute the linear recurrences for the~$c_{n,k}$ or~$C_{n,k}$ for given~$n$. First, the operators~$L_k$ can be computed as \emph{commutative} polynomials~$\tilde{L}_k$ as follows:
\begin{equation}\label{Ltilde}
\tilde{L}_{k+1}:=t\frac{\partial \tilde{L}_k}{\partial t}+\theta \tilde{L}_k-k(n-k+1)t^2\tilde{L}_{k-1},\qquad 1\le k\le n,
\end{equation}
with initial values~$\tilde{L}_0:=1$ and~$\tilde{L}_1:=\theta$.
These polynomials~$\tilde{L}_k$ coincide with the operators~$L_k$ when the powers of $\theta$ are written on the \emph{right} of the monomials in $t$ and $\theta$.

By collecting coefficients of~$t$ in~$\tilde{L}_{n+1}$, we recover~\eqref{finaldeq}.
Substituting $-1-k-j$ for $\theta$ in the coefficient of~$t^j$ then produces the desired recurrence for~$c_{n,k}$, while replacing~$c_{n,k+j}$ by $(k+1)\dotsm(k+j)C_{n,k+j}$ for all~$j$ produces one for~$C_{n,k}$.

\begin{ex} We illustrate the process for $n=4$.
The last operator in Example \ref{four} may be rewritten as
\[L_5=\theta^5-4t^2(5\theta^3+15\theta^2+18\theta+8)+64t^4(\theta+2)\]
and annihilates $K_0^4(t)$. Substituting $-1-k-j$ for $\theta$ in the coefficient of~$t^j$ for $j=0,2,4$  gives
\[-(k+1)^5c_{4,k}+4(k+2)(5k^2+20k+23)c_{4,k+2}-64(k+3)c_{4,k+4}=0.\]
Since $c_{4,k}=\frac{3}{2}\Gamma(k+1)C_{4,k}$, this is equivalent to
\[-\frac{3}{2}(k+1)^4C_{4,k}+6(k+2)^2(5k^2+20k+23)C_{4,k+2}-96(k+4)(k+3)^2(k+2)C_{4,k+4}=0\]
which was proven by different methods in~\cite{BaBoBoCr07}.
\end{ex}


Here is the corresponding \emph{Maple} code:
\begin{verbatim}
compute_Q:=proc(n,theta,t)
local k, L;
    L[0]:=1; L[1]:=theta;
    for k to n do
        L[k+1]:=expand(series(
            t*diff(L[k],t)+L[k]*theta-k*(n-k+1)*t^2*L[k-1],
            theta,infinity))
    od;
    series(convert(L[n+1],polynom),t,infinity)
end:

rec_c:=proc(c::name,n::posint,k::name)
local Q,theta,t,j;
    Q:=compute_Q(n,theta,t);
    add(factor(subs(theta=-1-k-j,coeff(Q,t,j)))*c(n,k+j),j=0..n+1)=0
end:

rec_C:=proc(C::name,n::posint,k::name)
local Q,theta,t,j,ell;
    Q:=compute_Q(n,theta,t);
    (-1)^(n+1)*(k+1)^n*C(n,k)+
        add(factor(subs(theta=-1-k-j,coeff(Q,t,j))
            *mul(k+1+ell,ell=1..j-1))*C(n,k+j),j=1..n+1)=0
end:
\end{verbatim}
On a reasonably recent personal computer, all recurrences for~$n$ up to~100 can be obtained in less than 5~minutes (further time could be saved by not factoring the coefficients). For example, the recursions for $c_{4,k}$ and~$C_{4,k}$ may be determined thus:
\begin{verbatim}
> rec_c(c, 4, k);
\end{verbatim}
\vskip-1ex
\[- (k+1)^5c_{4,k} +4 ( k+2 )  ( 5{k}^{2}+20k+23 ) c_{4,k+2} - (64k +192) c_{4,k+4} =0\]
\begin{verbatim}
> rec_C(C, 4, k);
\end{verbatim}
\vskip-2ex
\begin{multline*}
- ( k+1 ) ^{4}C_{{4,k}}+4 ( k+2 ) ^{2} ( 5{k}^{2}+20k+
23 ) C_{{4,k+2}}\\-64 ( k+4 )( k+3 ) ^{2} ( k+2 )   C_{{4,k+4}}=0
\end{multline*}

The first six cases for $C_{n,k}$ are
\begin{eqnarray}
0 &=& (k+1)C_{1,k} - (k+2)C_{1,k+2} \label{rec1} \\
0 &=& (k+1)^2 C_{2,k} - 4 (k+2)^2 C_{2,k+2} \label{rec2}\\
0 &=& (k+1)^3 C_{3,k}-2(k+2) \left(5(k+2)^2 + 1\right)C_{3,k+2}
 \nonumber \\
  && + 9 (k+2) (k+3) (k+4)C_{3,k+4} \label{rec3} \\
0 &=& (k + 1)^4 C_{4,k} - 4 (k+2)^2(5(k+2)^2 + 3) C_{4,k+2}
 \nonumber \\
  && + 64(k+2) (k + 3)^2(k + 4) C_{4,{k+4}} \label{rec4} \\
0 &{=}& (k+1)^5 C_{5,k}
  - (k+2) \left(35k^4+280 k^3+882 k^2+1288 k+731\right)C_{5,k+2}
  \nonumber \\
 && + (k+2) (k+3) (k+4) \left(259 k^2+1554 k+2435\right) C_{5,k+4}
  \nonumber \\
 && - 225 (k+2) (k+3) (k+4) (k+5) (k+6) C_{5,k+6} \label{rec5} \\
0 &{=}& (k+1)^6 C_{6,k}
 - 8 (k+2)^2 \left(7 k^4+56 k^3+182 k^2+280 k+171\right) C_{6,k+2}
  \nonumber \\
 && + 16 (k+2) (k+3)^2 (k+4) \left(49 k^2+294 k+500\right) C_{6,k+4}
  \nonumber \\
 && - 2304 (k+2) (k+3) (k+4)^2 (k+5) (k+6) C_{6,k+6}. \label{rec6}
\end{eqnarray}
as given in \cite{BaBoBoCr07}, but in which only the first four were proven (see also~\cite{ouvry} for an earlier proof up to $n=4$). Many more recursions were determined empirically using Integer Relation Methods---this relied on being able to compute the integrals in (\ref{integral}) to very high precision---and led to the  now-proven conjecture. The versions of these recurrences in terms of~$c_{n,k}$ instead of~$C_{n,k}$ were also determined empirically for~$n=1,\dots,6$ in~\cite[Eqs.~(11a--e)]{GuPr93} for the enumeration of staircase polygons.

Implicit in this algorithm is an explicit recursion for the polynomial coefficients of each recursion.  In the case of (\ref{rec3}) and (\ref{rec4}) these recursions lead to new continued fractions for $L_{-3}(2)$ and $\zeta(3)$ respectively~\cite{BaBoCr06, BaBoBoCr07}.
 These  rely additionally on the facts that $C_{3,1}=L_{-3}(2), C_{3,3}=2\,L_{-3}(2)/9 -4/27$ and
 $C_{4,1}=7\,\zeta(3)/12, C_{4,3}=7\,\zeta(3)/288-1/48,$   \cite{BaBoBoCr07}. Corresponding continued fractions arising from $C_{3,2}/C_{3,0}$ and  $C_{4,2}/C_{4,0}$  are determined in \cite{B3G08}.

\section{Another Example}
In~\cite{BaBoCr07} (to which we refer for motivation and references), the following ``box integrals'' have been considered
\begin{align*}
B_n(s)&=\int_0^1\dots\int_0^1(r_1^2+\dots+r_n^2)^{s/2}\,dr_1\dotsm dr_n,\\
\Delta_n(s)&=\int_0^1\dots\int_0^1((r_1-q_1)^2+\dots+(r_n-q_n)^2)^{s/2}\,dr_1\dotsm dr_ndq_1\dotsm dq_n.
\end{align*}
As in the case of the~$C_{n,k}$ we have considered here, a good starting point is provided by alternative integral representations for $\Re s>0$:
\begin{alignat*}{3}
B_n(-s)&=\frac{2}{\Gamma(s/2)}\int_0^\infty{u^{s-1}b(u)^n\,du},&\qquad b(u)&=\frac{\sqrt{\pi}\erf(u)}{2u}\\
\Delta_n(-s)&=\frac{2}{\Gamma(s/2)}\int_0^\infty{u^{-s-1}d(u)^n\,du},&\qquad d(u)&=\frac{e^{-u^2}-1+\sqrt{\pi}u\erf(u)}{u^2}.
\end{alignat*}
The first one is given explicitly as~\cite[(33)]{BaBoCr07} and the second one can be derived similarly. From classical properties of the error functions, the functions~$b(u)$ and~$d(u)$ satisfy the linear differential equations
\begin{align*}
ub''(u)+2(1+u^2)b'(u)+2ub(u)&=0,\\ 2u^2d'''(u)+4u(3+u^2)d''(u)+4(3+4u^2)d'(u)+8ud(u)&=0.
\end{align*}
This is exactly the set-up of our Proposition~\ref{stanley}. We thus deduce the existence of linear difference equations (wrt $s$) for both $B_n$ and $\Delta_n$. The fast computation of the difference equation for $B_n$ follows directly from the Algorithm of the previous section, and for instance, we get
\begin{multline*}
(s+9)(s+10)(s+11)(s+12)B_4(s+8)
-10(s+8)^2(s+9)(s+10)B_4(s+6)\\
+(s+6)(s+8)(35s^2+500s+1792)B_4(s+4)
-2(25s+148)(s+4)(s+6)^2B_4(s+2)\\
+24(s+2)(s+4)^2(s+6)B_4(s)=0.
\end{multline*}
The recurrence holds for all~$s$ by meromorphic continuation. A result on the shape of this recurrence for arbitrary~$n$ could be obtained along the lines of Lemma~3.

\medskip 

\noindent \textbf{Acknowledgements}  The authors wish to express their thanks to David Broadhurst for directing them to several relevant references and for his many incisive comments.

\bibliographystyle{plain}
\bibliography{ising}

\begin{thebibliography}{10}

\bibitem{AbSt73}
Milton Abramowitz and Irene~A. Stegun, editors.
\newblock {\em Handbook of mathematical functions with formulas, graphs, and
  mathematical tables}.
\newblock Dover Publications Inc., New York, 1992.
\newblock Reprint of the 1972 edition.

\bibitem{BaBoBoCr07}
D.~H. Bailey, D.~Borwein, J.~M. Borwein, and R.~E. Crandall.
\newblock Hypergeometric forms for {I}sing-class integrals.
\newblock {\em Experimental Mathematics}, 16(3):257--276, 2007.
\newblock \url{http://locutus.cs.dal.ca:8088/archive/00000326/}.

\bibitem{B3G08}
D.~H. Bailey, J.~M. Borwein, D.~M. Broadhurst, and L.~Glasser.
\newblock Elliptic integral representation of {B}essel moments.
\newblock {\em Journal of Physics. A.}, 2008.
\newblock To appear. \url{http://arxiv.org/abs/0801.0891}.

\bibitem{BaBoCr07}
D.~H. Bailey, J.~M. Borwein, and R.~E. Crandall.
\newblock Box integrals.
\newblock {\em Journal of Computational and Applied Mathematics},
  206(1):196--208, September 2007.

\bibitem{BaBoCr06}
D.~H. Bailey, J.~M. Borwein, and R.~E. Crandall.
\newblock Integrals of the {I}sing class.
\newblock {\em Journal of Physics. A.}, 39:12271--12302, 2007.

\bibitem{Broadhurst07}
David~J. Broadhurst.
\newblock Reciprocal {PSLQ} and the tiny nome of {B}ologna.
\newblock Talk at the Zentrum f{\"u}r interdisziplin{\"a}re Forschung in
  Bielefeld, June 2007.
\newblock
  \url{http://www.physik.uni-bielefeld.de/igs/schools/ZiF2007/Broadhurst.pdf}.

\bibitem{BrMuWe97}
Manuel Bronstein, Thom Mulders, and Jacques-Arthur Weil.
\newblock On symmetric powers of differential operators.
\newblock In Wolfgang~W. K{\"u}chlin, editor, {\em ISSAC '97}, pages 156--163,
  New York, NY, USA, 1997. ACM Press.

\bibitem{Chyzak00}
Fr{\'e}d{\'e}ric Chyzak.
\newblock An extension of {Z}eilberger's fast algorithm to general holonomic
  functions.
\newblock {\em Discrete Mathematics}, 217(1-3):115--134, 2000.

\bibitem{ChSa98}
Fr{\'e}d{\'e}ric Chyzak and Bruno Salvy.
\newblock Non-commutative elimination in {O}re algebras proves multivariate
  holonomic identities.
\newblock {\em Journal of Symbolic Computation}, 26(2):187--227, August 1998.

\bibitem{GuPr93}
A.~J. Guttmann and T.~Prellberg.
\newblock Staircase polygons, elliptic integrals, {H}eun functions, and lattice
  {G}reen functions.
\newblock {\em Physical Review E}, 47(4):2233--2236, April 1993.

\bibitem{Ince56}
E.~L. Ince.
\newblock {\em Ordinary differential equations}.
\newblock Dover Publications, New York, 1956.
\newblock Reprint of the 1926 edition.

\bibitem{ouvry}
St{\'e}phane Ouvry.
\newblock Random {A}haronov-{B}ohm vortices and some exactly solvable families
  of integrals.
\newblock {\em Journal of Statistical Mechanics: Theory and Experiment},
  1:P09004, 2005.
\newblock \url{http://arxiv.org/abs/cond-mat/0502366}.

\bibitem{SaZi94}
Bruno Salvy and Paul Zimmermann.
\newblock Gfun: a {M}aple package for the manipulation of generating and
  holonomic functions in one variable.
\newblock {\em ACM Transactions on Mathematical Software}, 20(2):163--177,
  1994.

\bibitem{Stanley99}
Richard~P. Stanley.
\newblock {\em Enumerative combinatorics}, volume~2.
\newblock Cambridge University Press, 1999.

\bibitem{vdPSi02}
Marius van~der Put and Michael~F. Singer.
\newblock {\em Galois theory of linear differential equations}, volume 328 of
  {\em Grundlehren der Mathematischen Wissenschaften [Fundamental Principles of
  Mathematical Sciences]}.
\newblock Springer-Verlag, Berlin, 2nd edition, 2002.

\bibitem{Zeilberger90}
Doron Zeilberger.
\newblock A holonomic systems approach to special functions identities.
\newblock {\em Journal of Computational and Applied Mathematics},
  32(3):321--368, 1990.

\end{thebibliography}
\end{document}